\documentclass[aps,preprint]{revtex4}

\usepackage{graphics,graphicx}
\usepackage{amsmath,amssymb}
\usepackage{epsfig}
\usepackage{txfonts}
\usepackage{subfigure}
\usepackage[dvipsnames,svgnames]{xcolor}
\usepackage{epstopdf}
\usepackage[colorlinks=true,linkcolor=DarkRed,urlcolor=DarkBlue,citecolor=DarkGreen]{hyperref}
\usepackage[sort&compress]{natbib}
\begin{document}

\title{Kalb-Ramond field localization on a de Sitter Thick Brane}
\author{Chen Yang $^{1,2,}$\footnote{Email: yangch2017@lzu.edu.cn}}
\author{Zi-Qi Chen$^{1,2,}$\footnote{Email: chenzq16@lzu.edu.cn}}
\author{Li Zhao $^{1,2}$\footnote{Email: lizhao@lzu.edu.cn, corresponding author}}

\affiliation{\small{$^{1}$ Institute of Theoretical Physics, Lanzhou University, 222 South Tianshui Road, Lanzhou 730000, China\\
                    $^{2}$ Research Center of Gravitation, Lanzhou University,  222 South Tianshui Road, Lanzhou 730000, China\\
                  }}

\begin{abstract}
In this paper, we study the localization of Kalb-Ramond (KR) tensorial gauge field on a non-flat de Sitter thick brane. The localization and resonance of KR gauge field are discussed for three kinds of couplings.  For the first coupling there is no localized tensorial zero mode. For the other two couplings, the zero mode of KR field  can be localized under certain condition.  There are resonant KK
modes on the thick brane for the third case.  Furthermore, we mainly analyze the effects of three parameters on the localization and resonant mode  for KR field.
\end{abstract}


\maketitle

\section{Introduction}
Inspired by the String/M theory, the braneworld theory  was proposed as an alternative to solve the  long-existing  gauge hierarchy problem \cite{Randall1999,Guo2015b,Yang2012,Antoniadis1998,Arkani-Hamed1998,Gogberashvili2002} and  cosmology problem \cite{Arkani-Hamed2000,Neupane2011,Starkman2001,Kim2001}. This theory has attracted much  attention since  Arkani-Hamed, Dimopoulos and Dvali (ADD) \cite{Arkani-Hamed1998} and Randall-Sundrum (RS) (resp.RS1, RS2) models \cite{Randall1999,Randall1999b} were presented, with a new idea that we live on  a thin membrane embedded into a higher-dimensional bulk. In the original RS models, the  brane thickness is ignored. However, from string theory's viewpoint, a brane should have a minimal length scale.  In the field theories, the scalar fields are introduced to generate the topological defects,  which are essential for the appearance of  thick brane \cite{Rubakov:1983rsd,Chumbes2012}.

The localization of various matter fields is an interesting
subject in braneworld senario, because it guides us to understand  the zero mode of bulk fields more phenomenologically. In general, massless scalar field and
graviton can be localized in the RS thin brane or its generalizations \cite{BajcPLB2000,Liu0708,Koroteev08,Flachi09},
but it is not true for free vector and Kalb-Ramond (KR) tensor fields \cite{Neronov2001,Cruz2009,Tahim2009,Christiansen2012}.
The localization of the fermion field  can be guaranteed by
introducing the scalar-fermion coupling \cite{ThickBrane2,LiuXu2014}, or a derivative geometrical coupling \cite{LiYY2017}.
While for a free massless vector field, it cannot be localized
on the RS2-like branes in five dimensions \cite{Davoudias2000}, but  can be localized on some non-RS2 brane models, for instance, the thick  Weyl brane  \cite{Liu2008} and  a string-like defect \cite{Oda2000} and so on. In addition, the vector field can  be localized on the thick brane  if one adds a dilaton scalar field \cite{Kehagias2001}, a 5D Stueckelberg-like action \cite{Vaquera-Araujo2015} or  a topology term to the 5D vector action \cite{Odd2001}.  Recently, one of work is of particular interest: the authors
of \cite{Zhaozh2017} suggest a localization mechanism which includes a function of scalar curvature in the 5D vector field action.

In low-energy superstring models, the effective theories support the existence of higher spin fields.  Motivated by the previous researches on vector field, our main subject is to answer the following question: how to localize the higher-spin gauge field on a thick brane? The simplest case  of higher rank fields to be studied is the KR field (i.e., a rank-2 antisymmetric field). KR field  emerges as a mode of massless excitations  of closed strings \cite{Green1985}, and  describes the torsion of a Riemannian-Cartan manifold \cite{Vasilic2008} as well. For the free KR field, it is  failed to localize it in a thick brane with the standard action.   The mechanisms for gauge field localization are mainly three-fold: firstly, a general mechanism with a coupling between the gauge field and an additional dilaton field \cite{Kehagias2001,Moazzen2017,Cruz2013,Christiansen2012}; secondly, a mechanism to localize gauge field on thick brane directly coupled to the background scalar field \cite{Chumbes2012}; thirdly, the mechanism for gauge field localization on thick branes based on a five-dimensional Stueckelberg-like action \cite{Vaquera-Araujo2015}.

 The previous works on the  localization of the  Kalb-Ramond field are  investigated based  on a flat 3-brane \cite{Cruz2009,Christiansen2010,Cruz2013}. In this paper,  following the work of Ref. \cite{Chumbes2012} with the coupling directly with the background kink scalar, we will analyze the localization on a non-flat dS 3-brane. In more detail, the brane possesses de Sitter symmetry instead of the Poincar\'{e} one. For positive scale factor parameter, the brane has a  expanding solution. While for the  vanishing parameter,  the brane has a Minkowski metric.

In Sec. \ref{sec:model}, we briefly review  the action of the thick brane system and derive the equations of
motion. Then, in Sec. \ref{sec:perturbation}, a Schr\"{o}dinger-like equation  and the corresponding effective potential are obtained.
 In Sec. \ref{sec:localization2},  we investigate the localization and resonance of KR tensor fields on this thick brane.
 Finally, our conclusion is given in Sec. \ref{conclusion}.

\section{Brane Setup and field equations}
\label{sec:model}
We consider a thick dS brane embedded into a 5D spacetime, and
the thick  dS brane is generated by one real scalar field  coupled to
gravity minimally. The action of our system is expressed as
\begin{align}
S=&\int\mathrm{d}^5x\sqrt{-g}\left[\frac{1}{2\kappa_5^2}R-\frac{1}{2}g^{MN}
\partial_{M}\phi\partial_{N}\phi
-U(\phi)\right],\label{action}
\end{align}
where $\kappa_5$ is the 5D coupling constant, and $U(\phi)$ is  the scalar field potential. Here
we set $\kappa_5=\hbar=c=1$ for simplicity.

\begin{figure}[htb]
\centering
\includegraphics[width=2.6in]{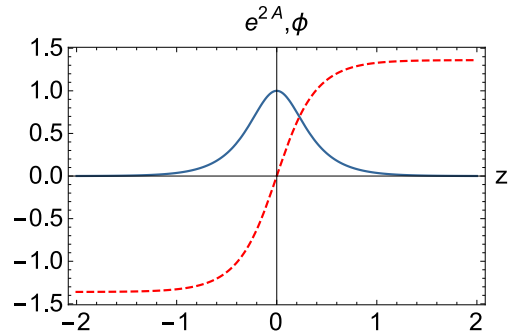}}
\hspace{0.5in}
\caption{Plots of $e^{2A(z)}$ and $\phi(z)$, where $e^{2A(z)}$ is the warp factor in Eq.~\eqref{e2A1} and $\phi(z)$ is the scalar field in Eq.~\eqref{phi1}. The parameters are set to $\delta=1/2,\beta=2$.}\label{fig:ay&phiy}
{\label{subfig:ay1}
\end{figure}

\begin{figure}[htb]
\centering
\includegraphics[width=2.8in]{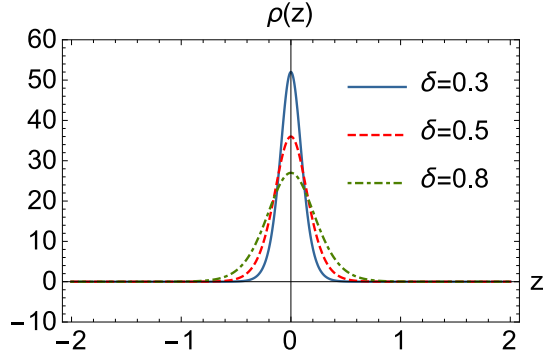}}
\hspace{0.5in}
\caption{Plots of the energy density $\rho(z)$ in Eq.~\eqref{EnergyDensity}. The parameter is set to $\beta=2$.}
{\label{subfig:ay2}
\end{figure}

For a expanding and  maximally symmetric dS brane, the 5D metric ansatz is assumed as
\begin{eqnarray}
 ds^2&=&\text{e}^{2A(z)}\big(\hat{g}_{\mu\nu}dx^\mu dx^\nu
          + dz^2\big)    \nonumber \\
 &=&\text{e}
 ^{2A(z)}\big(-dt^2+e^{2\beta t}dx^i dx^i + dz^2\big),
\label{linee1}
\end{eqnarray}
where $e^{2A(z)}$ is the warp factor and and $e^{2\beta t}$ is the scale factor of the brane. For positive and vanishing  scale factor parameter $\beta$, one will obtain the expanding and static solutions, respectively. The warp factor $A(z)$ and the scalar field $\phi$
only depend on the extra coordinate $z$. With the metric ansatz~\eqref{linee1}, the variation of the  action~\eqref{action} with respect to the metric $g_{MN}$ and the scalar field $\phi$ yields the following field equations,
\begin{eqnarray}
\phi'^2 & = & 3(A'^2-A''-\beta^2), \label{coupledEqa} \\
U(\phi) & = & \frac{3}{2} e^{-2A}
 (-3A'^2-A''+3\beta^2), \label{coupledEqb}\\
\frac{dU(\phi)}{d\phi} &  = & e^{-2A}(3A'\phi'+\phi''),
\label{coupledEqc}
\end{eqnarray}
where the prime denotes derivative with respect to $z$.
A thick  dS brane solution  for this system was found in  Refs. \cite{Goetz1990,Gass1999}:
\begin{eqnarray}
U(\phi)&=&\frac{1+3\delta}{2\delta}\ 3\beta^{2}\left(\cos
\frac{\phi}{\phi_{0}} \right)^{2(1-\delta)}
          \label{potencial goetz}\\
 e^{2A}&=&\cosh^{-2\delta}\left(\frac{\beta z}{\delta}\right) ,
         \label{e2A1} \\
 \phi~&=&\phi_{0}\arctan\left(\sinh \frac{\beta z}{\delta}\right),
         \label{phi1}
\end{eqnarray}
where $\phi_{0} =\sqrt{3\delta(1-\delta)}$, $0<\delta<1, \beta>0$.
The warp factor and the background scalar field are depicted in Fig. (\ref{subfig:ay1}).
It can be seen that the warp factor indicates a thick brane centered at the origin $y = 0$,
and the scalar field has a kink configuration with $\pm\frac{\pi}{2}\phi_{0}$ at $z\rightarrow\pm
\infty$, corresponding to two consecutive minima of the potential $U(\phi)$.
From the expressions of warp factor and  scalar field, one can calculate the energy density $\rho(z)$ along the extra dimension,
\begin{eqnarray}
 \rho(z) &=&\frac{3\beta^2(1+\delta)}{\delta}
  \cosh^{-2-2\delta}\left(\frac{\beta z}{\delta}\right).
       \label{EnergyDensity}
\end{eqnarray}
which is plotted in Fig.(\ref{subfig:ay2}).   This figure shows that increasing the values of
the parameter $\delta$ makes the brane thicker, and $\delta$ parameterizes the thickness of the 3-brane.

\section{Schr\"{o}dinger-like equation and effective potential}
\label{sec:perturbation}
In this part, we will discuss the influence of coupling with the background scalar field on the localization of KR field. The KR field $B_{\mu\nu}$ couples directly with background scalar field in a combination defined by the action:
\begin{equation}\label{perturbation}
S=\int d^{5}x \sqrt{-g}\left[f(\phi)H_{MNL}H^{MNL}\right],
\end{equation}
where $H_{MNL}=\partial_{[M}B_{NL]}$ is the field strength, $f(\phi)$ is the coupling with the background scalar field.
From the action and metric ansatz, the equation of motion for the KR  field is expressed as
\begin{equation}
\partial_{R}\left[\sqrt{-g}f(\phi)H^{RPQ}\right]=0.
\end{equation}
Considering the antisymmetry of the KR field strength, the equation of motion can be transformed into
\begin{eqnarray}\label{fphi}
f(\phi)\partial_{\mu}\left[\sqrt{-g}H^{\mu\nu\lambda}\right]+\partial_{5}\left[\sqrt{-g}f(\phi)H^{5\nu\lambda}\right]=0, \nonumber \\
f(\phi)\partial_{\mu}\left[\sqrt{-g}H^{\mu\nu5}\right]=0.
\end{eqnarray}
By choosing the gauge condition $B_{\mu5}=0$, and making the Kaluza-Klein (KK) decompositon of the KR field as
\begin{equation}
B^{\nu\lambda}=\sum_{n} b^{\nu\lambda}_{(n)}(x)\chi_{(n)}(z)\Omega(z),\label{KKdecompositon}
\end{equation}
the field strength becomes
\begin{equation}
H_{\mu\nu\lambda}=\sum_{n} e^{4A(z)} h_{\mu\nu\lambda(n)}(x)\chi_{(n)}(z)\Omega(z), \label{fieldstrength}
\end{equation}
where $h_{\mu\nu\lambda(n)}(x)$  is the 4D part of KR field  strength on the brane.
Substituting Eq.(\ref{KKdecompositon}) into Eqs.(\ref{fphi}), we get the following equations
\begin{eqnarray}
\partial_{\tau}\left(\sqrt{-\hat{g}}h^{\tau\mu\nu}_{(n)}\right)&=&\frac{1}{3}m_{n}^{2}b^{\mu\nu}_{(n)}\sqrt{-\hat{g}},\\
\partial_{5}\left(e^{-A(z)}f(\phi)\partial_{5}\left(e^{4A(z)}\chi_{(n)}(z)\right)\Omega(z)\right)&=&-m_{n}^{2}e^{3A(z)}f(\phi)\chi_{(n)}(z)\Omega(z). \label{fieldstrength2}
\end{eqnarray}
By eliminating the first derivative term of $\chi_{(n)}(z)$, i.e.,
\begin{equation}
7A'(z)f(\phi)\Omega(z)+f'(\phi)\Omega(z)+2f(\phi)\Omega'(z)=0,\label{firstderivative}
\end{equation}
Eq.(\ref{fieldstrength2}) can be  transformed into a Schr\"{o}dinger-like equation,
\begin{equation}
\left[-\frac{d^{2}}{dz^{2}}+V_{\text{eff}}(z)\right]\chi_{(n)}(z)=m_{n}^{2}\chi_{(n)}(z), \label{Schrodingerequation}
\end{equation}
where the effective potential $V_{\text{eff}}(z)$ of the Schr\"{o}dinger-like equation is given by
\begin{eqnarray}
V_\text{eff}(z)&=&\left(4 A'(z)+\frac{\Omega'(z)}{\Omega(z)}\right)^2-\left(4 A'(z)+\frac{\Omega'(z)}{\Omega(z)}\right)'.
 \label{effectvepotential}
\end{eqnarray}
 In fact, the Schr\"{o}dinger-like equation  (\ref{Schrodingerequation}) can also be written as
 \begin{equation}
\mathcal{H}\chi_{(n)}(z)=m_{n}^{2}\chi_{(n)}(z),\label{eq:Schrodinger}
\end{equation}
 where
\begin{equation}
\mathcal{H}=Q^{+}Q=\left(-\partial_{z}+4A'(z)+\frac{\Omega'(z)}{\Omega(z)}\right)\left(\partial_{z}+4A'(z)+\frac{\Omega'(z)}{\Omega(z)}\right)
\end{equation}
with $Q=\partial_{z}+4A'(z)+\frac{\Omega'(z)}{\Omega(z)}$. As the operator $\mathcal{H}$ is positive definite, there are no tachyonic
 modes with negative $m_{n}^{2}$.
Considering the KK decomposition of the KR field (\ref{KKdecompositon}), we derive the 5D action  as
\begin{eqnarray}
S_{\text{KR}}&=& \int d^{4}x \sqrt{-\hat{g}}
\left[f(\phi)h_{\mu\nu\lambda(l)}h^{\mu\nu\lambda}_{(m)}+\frac{1}{3}m^{2}_{l}\sqrt{-\hat{g}}b_{\mu\lambda(l)}(x)b^{\mu\lambda}_{(m)}\right]
\int e^{7A(z)}f(\phi)\chi_{(l)}(z)\chi_{(m)}(z)\Omega^{2}(z)dz.  \nonumber
\end{eqnarray}
It is worth to note that if we impose the orthonormality condition between different massive modes,
\begin{equation}
\int e^{7A(z)}f(\phi)\chi_{(l)}(z)\chi_{(m)}(z)\Omega^{2}(z)dz=\delta_{lm},
\end{equation}
the 5D action can be reduced to the 4D effective one,
\begin{eqnarray}
S_{\text{KR}}\propto S_{\text{eff}}=\int d^{4}x \sqrt{-\hat{g}}
\left[f(\phi)h_{\mu\nu\lambda(m)}h^{\mu\nu\lambda}_{(m)}+\frac{1}{3}m^{2}_{m}\sqrt{-\hat{g}}b_{\mu\lambda(m)}(x)b^{\mu\lambda}_{(m)}\right].
\nonumber
\end{eqnarray}

\section{Massive spectra and resonance}
\label{sec:localization2}
In this section we will investigate the localization of KR field under three different couplings directly with the background scalar field.
If the KR field is localized with certain kind of coupling, we  further investigate the  mass spectra  of the  KK mode or resonant structure.

\subsection{Case I: \texorpdfstring{$f(\phi)=\phi^{k}$}{f(\phi)=\phi^{k}}}
\label{caseI}

Firstly, we consider the simple case of $f(\phi)=\phi^{k}$, where $\phi$ is the  background scalar field and $k$ is a positive parameter. From Eq. (\ref{firstderivative}), the function $\Omega(z)$ for this case reads
\begin{equation}
\Omega(z)=e^{-\frac{7}{2}A(z)}f(\phi)^{-\frac{1}{2}}= \cosh ^{\frac{7 \delta}{2}}\left(\frac{\beta z}{\delta}\right) \left[\sqrt{3(1-\delta) \delta} \tan ^{-1}\left(\sinh \left(\frac{\beta z}{\delta}\right)\right)\right]^{-\frac{k}{2}}.
\end{equation}
According to Eq.(\ref{effectvepotential}), the corresponding effective potential is expressed as
\begin{eqnarray}
V_{\text{eff}}(z)&=&\frac{\beta ^2 \text{sech}^2\left(\frac{\beta  z}{\delta }\right)}{8 \delta ^2 \tan ^{-1}\left(\sinh \left(\frac{\beta  z}{\delta }\right)\right)^2}[2 k^2+4 (\delta -1) k \sinh \left(\frac{\beta  z}{\delta }\right) \tan ^{-1}\left(\sinh \left(\frac{\beta  z}{\delta }\right)\right)-4 k\\ &&-\delta ^2 \tan ^{-1}\left(\sinh \left(\frac{\beta  z}{\delta }\right)\right)^2+\delta ^2 \cosh \left(\frac{2 \beta  z}{\delta }\right) \tan ^{-1}\left(\sinh \left(\frac{\beta  z}{\delta }\right)\right)^2+4 \delta  \tan ^{-1}\left(\sinh \left(\frac{\beta  z}{\delta }\right)\right)^2].
\end{eqnarray}
The values of $V_{\text{eff}}(z)$ at $z=0$ and $z\rightarrow\pm \infty$ read
\begin{eqnarray}
    &&V_{\text{eff}}(0)=\left\{
                \begin{aligned}
                 +\infty,~ 0<k<2 \\
                 -\infty,~k>2\\
                \end{aligned}
                \right. \nonumber \\
    &&V_{\text{eff}}(\pm \infty)=0.
\end{eqnarray}
We can find the effective potential tends to be zero when $z\rightarrow\pm \infty$.  In the case of $0<k<2$,  the values of $V_{\text{eff}}(z)\rightarrow+\infty$ when $z\rightarrow 0$, and the potential has a potential barrier.  Otherwise, $V_{\text{eff}}(z)\rightarrow-\infty$ when  $z\rightarrow 0$ , and the potential has a potential well.
From the Schr\"{o}dinger-like equation~\eqref{eq:Schrodinger}, the solutions of the KR field zero modes are given by
\begin{equation}\label{zeromode}
\chi_{(0)}(z)\propto e^{-4A(z)}\Omega(z)^{-1}.
\end{equation}
In order to localize the zero modes on the brane, they should satisfy the normalization
condition
\begin{equation}
\int_{-\infty}^{+\infty} e^{7A(z)}f(\phi)\chi_{(0)}(z)\chi_{(0)}(z)\Omega^{2}(z)dz<\infty,
\end{equation}
which can be converted into
\begin{equation}
\int_{-\infty}^{+\infty} e^{-8A(z)}\Omega^{-2}(z)dz<\infty. \label{integral}
\end{equation}
The integrand function is expressed as
\begin{eqnarray}
e^{-8A(z)}\Omega^{-2}(z)
 = \mathrm{sech}^{-\delta}\left(\frac{\beta z}{\delta}\right) \left[\sqrt{-3 \delta(\delta-1)}\tan^{-1}\left(\sinh \left(\frac{\beta z}{\delta}\right)\right)\right]^{k}.
\end{eqnarray}
When $z\rightarrow \infty$, $\mathrm{sech}(\frac{\beta z}{\delta})^{-\delta}\rightarrow 0$ and $\tan^{-k}(\sinh(\frac{\beta z}{\delta}))\rightarrow (\frac{\pi}{2})^{k}$ with $\delta< 0$.  The integral in Eq. (\ref{integral}) is finite when  $\delta< 0$, however, the background solution requires $0<\delta<1$. Thus, the value  of $\delta$ does not satisfy the normalization condition and this coupling fails to make the KR field localize on the thick dS brane.

\subsection{Case II: \texorpdfstring{$f(\phi)=\cos^{k}\phi$}{f(\phi)=\cos^{k}\phi}}
\begin{figure}[htb]
\begin{center}
\subfigure[$\delta=0.2, k=3$]{
\includegraphics[width=2.6in]{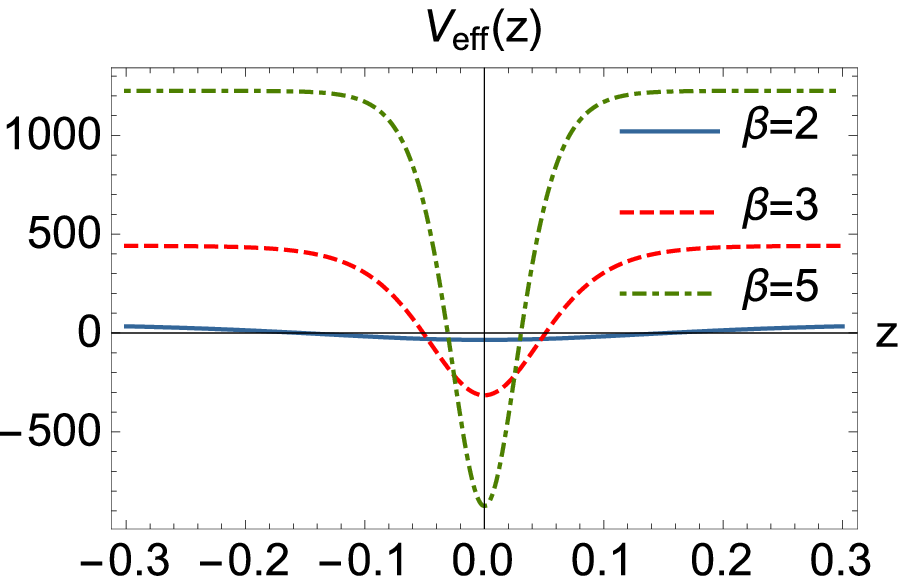}}
\subfigure[$\beta=3, k=3$]{
\includegraphics[width=2.6in]{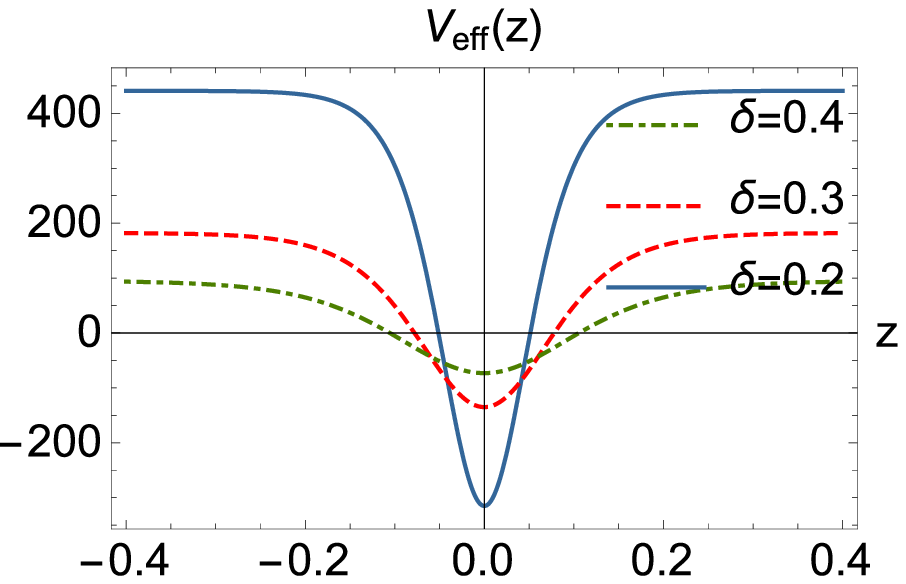}}
\subfigure[$\beta=3, \delta=0.2$]{
\includegraphics[width=2.6in]{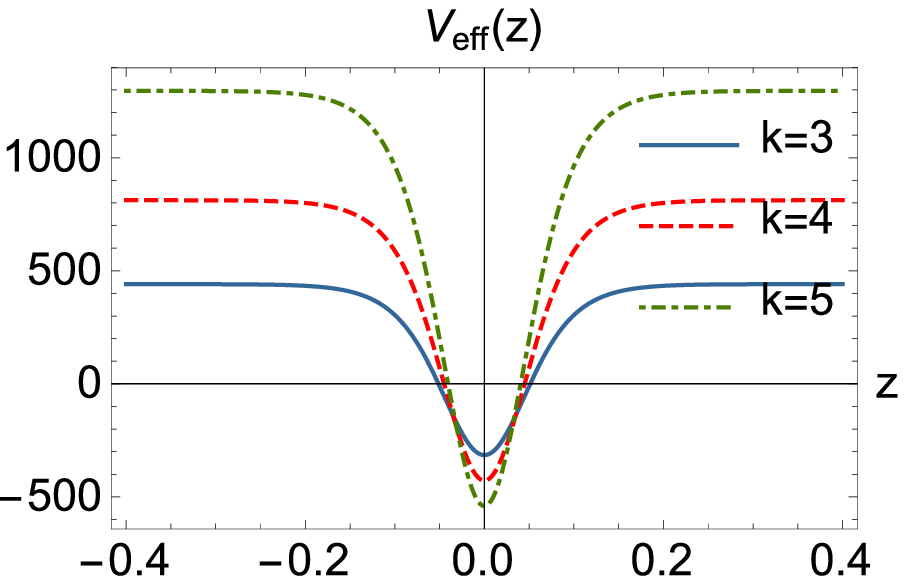}}
\end{center}
\caption{
Plots of the effective potentials $V_{\text{eff}}(z)$  for  different values of $\beta$, $\delta$ and $k$.
}
\label{VLVR2-1}
\end{figure}

Next we consider $f(\phi)=\cos^{k}\phi$,  for this coupling the effective potential is given by
\begin{eqnarray}
V_{\text{eff}}(z)&=&\frac{\beta ^2 (\delta -k) \text{sech}^2\left(\frac{\beta  z}{\delta }\right) \left(-\delta +(\delta -k) \cosh \left(\frac{2 \beta  z}{\delta }\right)+k+4\right)}{8 \delta ^2},\label{effective potential}
\end{eqnarray}
which has the asymptotic behavior of a finite square-well like
potential: $V_{\text{eff}}(z\rightarrow\pm \infty)\rightarrow constant$ and $V_{\text{eff}}(0)=\frac{\beta ^2 (\delta -k)}{2 \delta ^2}$. Since $V_{\text{eff}}(0)<0$ with $\delta <k$ and the value of
$V_{\text{eff}}(z\rightarrow 0)$ is a positive constant at infinity  (this kind of potential is also called P\"{o}schl-Teller (PT)-like), the effective potential provides a mass
gap to separate the zero mode from the KK modes.
Figure ~(\ref{VLVR2-1}) gives a detailed
description of the effect of the parameters $\beta$, $\delta$ and $k$ on
the  effective potentials. It can be seen that, with the increase of $\beta$ and $k$, or the decrease of $\delta$,  the
effective potential becomes deeper and thinner.

\begin{figure}[htb]
\begin{center}
\subfigure[$\beta=0.2, \delta=0.2, k=3$]{
\includegraphics[width=2.6in]{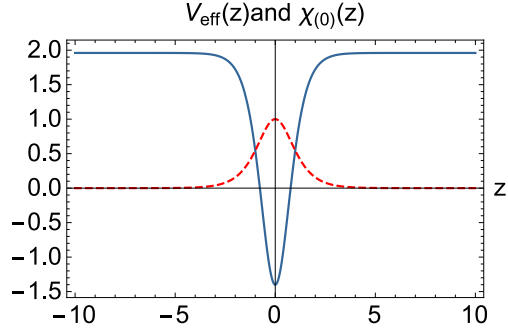}}
\end{center}
\caption{
Plots of the effective potentials $V_{\text{eff}}(z)$ (the blue line) and the zero
mode $\chi_{(0)}(z)$ (the dashed red line)  for  coupling
II. The parameters are set to $\beta=0.2, \delta=0.2, k=3$.}
\label{Veffwavefunction}
\end{figure}

The expression of the zero mode $\chi_{(0)}(z)$ in this case reads
\begin{equation}
\chi_{(0)}(z)\propto\cosh^{\frac{\delta-k}{2}} \left(\frac{\beta z}{\delta}\right),
\end{equation}
and  the shapes of the effective potential $V_{\text{eff}}$ and the zero
mode $\chi_{0}(z)$ are shown in Fig. ~(\ref{Veffwavefunction}).
As the case I, the normalization condition is
\begin{equation}
\int \chi_{(0)}^{2}(z)dz<\infty.
\end{equation}
When $z\rightarrow \infty$, $\cosh^{\delta-k}(\frac{\beta z}{\delta})\rightarrow 0$ with the condition $\delta-k< 0$.  We have known the background solution requires $0<\delta<1$. Therefore, in order to localize the KR field on the brane, we require the parameter $k$ satisfies the condition $k > \delta\in (0,1)$.

\begin{figure}[htb]
\begin{center}
\subfigure[$\alpha =8, h = 2$]{
\includegraphics[width=2.6in]{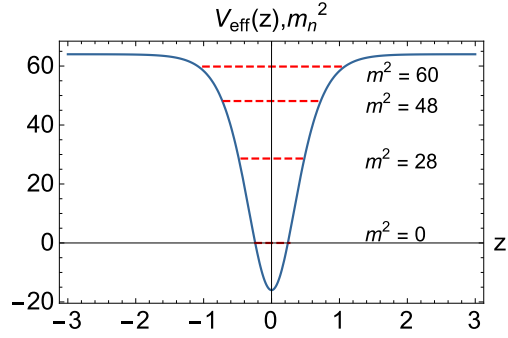}}
\end{center}
\caption{Plots of the effective potential of scalar KK modes
and the mass spectra (dashed red lines) for KR field with the parameter $\alpha =8, h = 2$.}
\label{massspectrum}
\end{figure}

\begin{figure}[htb]
\begin{center}
\subfigure[$n=0$]{
\includegraphics[width=2.6in]{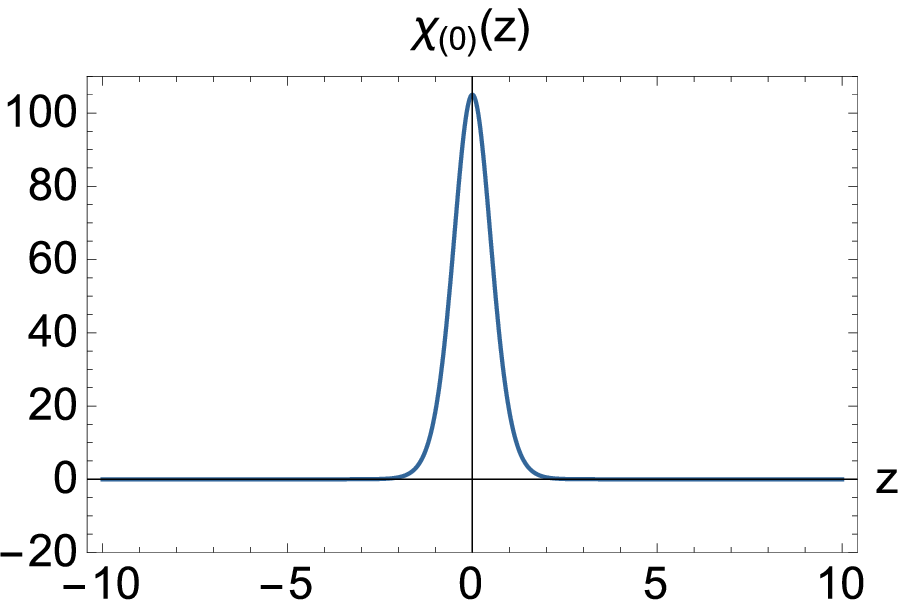}}
\subfigure[$n=1$]{
\includegraphics[width=2.6in]{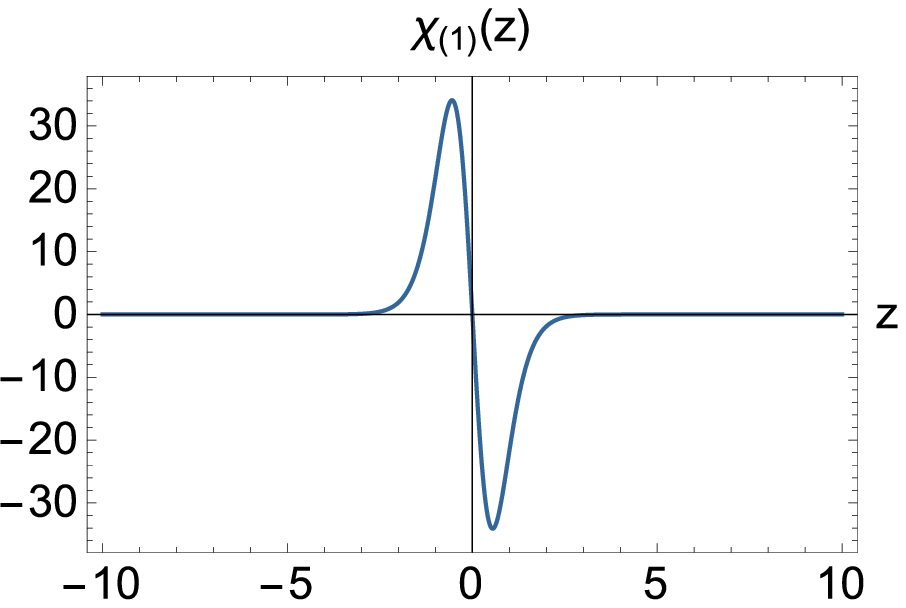}}\\
\subfigure[$n=2$]{
\includegraphics[width=2.6in]{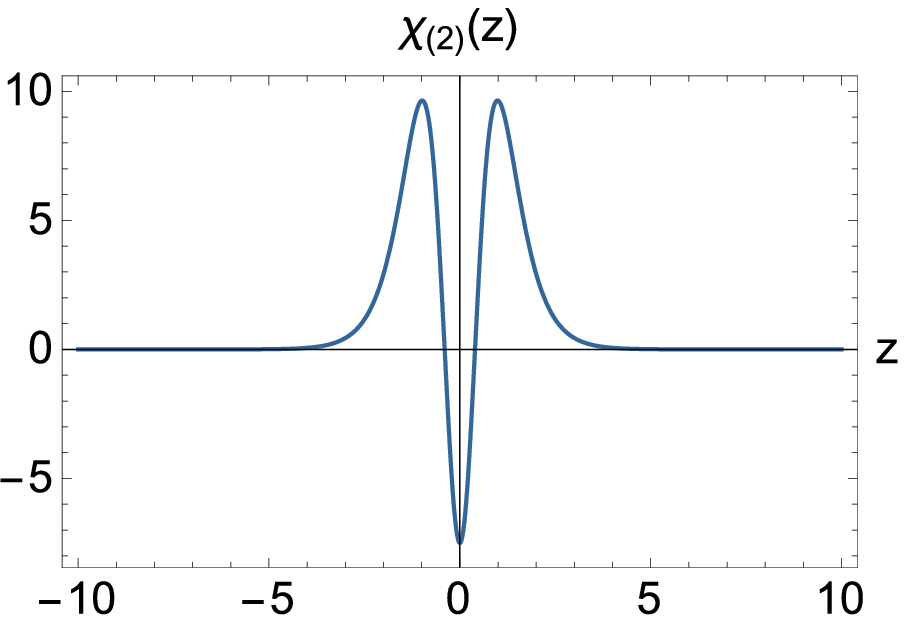}}
\subfigure[$n=3$]{
\includegraphics[width=2.6in]{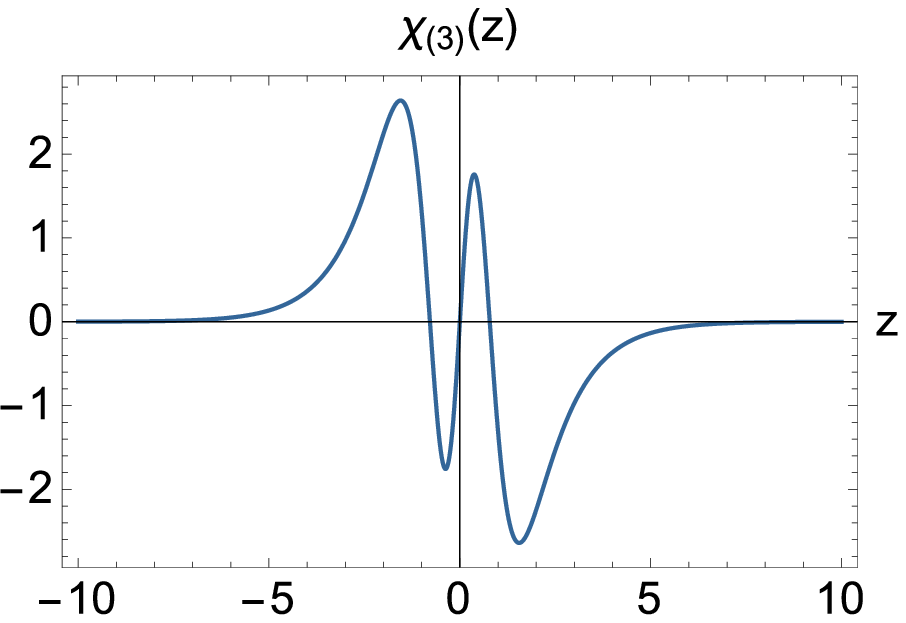}}
\end{center}
\caption{
Plots of the bound KK modes for KR field. The parameters are set to $\alpha=8$ and $h=2$.}
\label{PTwavefunction}
\end{figure}

 Next, by introducing the parameters $h=\beta/\delta, \alpha=k-\beta$, the Schr\"{o}dinger-like equation in Eq.(\ref{Schrodingerequation}) is derived as:
\begin{equation}
\left[-\partial_{z}^{2}-\left(\frac{\alpha}{2}+\frac{\alpha^{2}}{4}\right)h^{2}\text{sech}^{2}(hz)\right]\chi_{(n)}(z)=E_{n}\chi_{(n)}(z), \label{schrodingerlikeequation}
\end{equation}
where the energy is  $E_{n}=m_n^2-\alpha^{2}h^2/4$. If  we further perform the following
rescaling of the fifth coordinate $v =hz$, and suppose $\tau= \alpha/2+\alpha^2/4 $,  Eq. (\ref{schrodingerlikeequation}) can be written in the canonical form:
\begin{equation}
\left[-\partial_{v}^{2}-\tau^{2}\text{sech}^{2}(v)\right]\chi_{(n)}(z)=E_{n}\chi_{(n)}(z),\label{canonical form}
\end{equation}
where the energy now reads $E_{n}=m_n^2/h^{2} -\alpha^2/4$.
If the Schr\"{o}dinger equation has a modified PT potential, it has a exactly energy spectrum of bound states with $\tau=n(n+1)$, where $n$ is the number of bound states. In this case the energy spectrum  of bound states reads
\begin{equation}
E_{n}=-h^2\left(\frac{\alpha}{2}-n\right)^{2},
\end{equation}
or, in terms of the squared mass:
\begin{equation}
m_{n}^{2}=n(\alpha-n)h^{2}.
\end{equation}
Changing $\alpha$ and $h$, we can get a series of KK mode massive spectrum.
For example, the mass spectra $m_{n}^{2}$ of the KK mode are listed as:
\begin{eqnarray}
     m_{n}^2 &=& (0,3),
     ~\text{for}~\alpha =4, h = 1;\\
     m_{n}^2 &=& (0,5,8),
     ~\text{for}~\alpha =6, h = 1;\\
     m_{n}^2 &=& (0,7,12,15),~\text{for}~\alpha =8, h = 1;\\
     m_{n}^2 &=& (0,28,48,60),
     ~\text{for}~\alpha =8, h = 2;\\
     m_{n}^2 &=& (0,63,108,135),~\text{for}~\alpha =8, h = 3.
  \end{eqnarray}
The shapes of the effective potential and the mass spectra are shown Fig.(\ref{massspectrum}), which reveals the massive KK modes asymptotically turn
into continuous plane waves when $m^{2}>60$.
By solving Eq.(\ref{schrodingerlikeequation}), we get the general solution of massive KK modes as
\begin{equation}
\chi_{(n)}= C_1 P_{\frac{\alpha}{2}}^{u}(\tanh (hz))+C_2 Q_{\frac{\alpha}{2}}^{u}(\tanh (hz))
\end{equation}
where $C_1$ and $C_2$ are arbitrary constants and $u= \sqrt{\frac{a^2}{4}-\frac{ m_{n}^2}{h^2}}$; $P_{\frac{\alpha}{2}}^{u}$ and $Q_{\frac{\alpha}{2}}^{u}$ are associated Legendre functions
of first and second kind, respectively. The shapes of  wave function of four bound KK modes are shown in Fig. (\ref{PTwavefunction})  for the parameters $\alpha=8$ and $h=2$.

\subsection{Case III: \texorpdfstring{$f(\phi)=\left(\sinh ^2\left(\frac{\beta z}{\delta}\right)+1\right)^{-\frac{\delta}{2}}/\left(\sinh ^{-1}\left(\sinh \left(\frac{\beta z}{\delta}\right)\right)^2+\frac{1}{\delta^2}\right)^{-k}$}{f(\phi)=\left(\sinh ^2\left(\frac{\beta z}{\delta}\right)+1\right)^{-\frac{\delta}{2}}/\left(\sinh ^{-1}\left(\sinh \left(\frac{\beta z}{\delta}\right)\right)^2+\frac{1}{\delta^2}\right)^{-k}}}

\begin{figure}[htb]
\begin{center}
\subfigure[$\beta=2$]{
\includegraphics[width=2.6in]{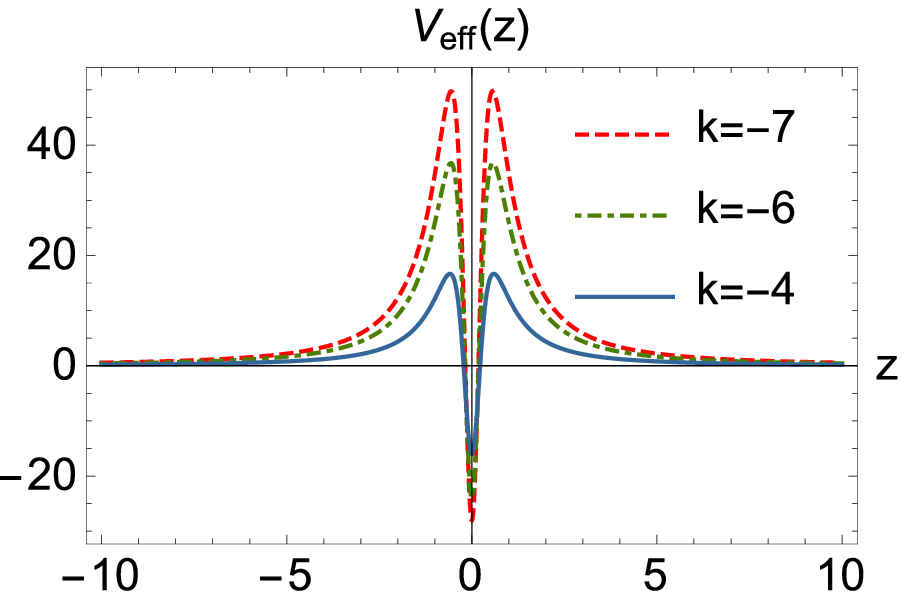}}
\subfigure[$k=-7$]{
\includegraphics[width=2.6in]{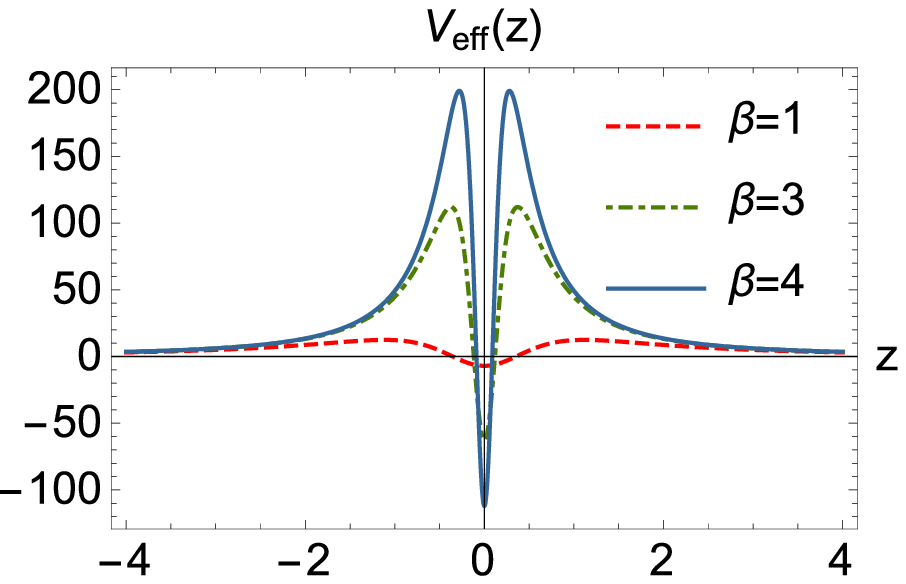}}
\end{center}
\caption{
Plots of the effective potential for KR field with different parameters $k$ and $\beta$.}
\label{Volcanowavefunction}
\end{figure}

\begin{figure}[htb]
\begin{center}
\subfigure[]{
\includegraphics[width=2.6in]{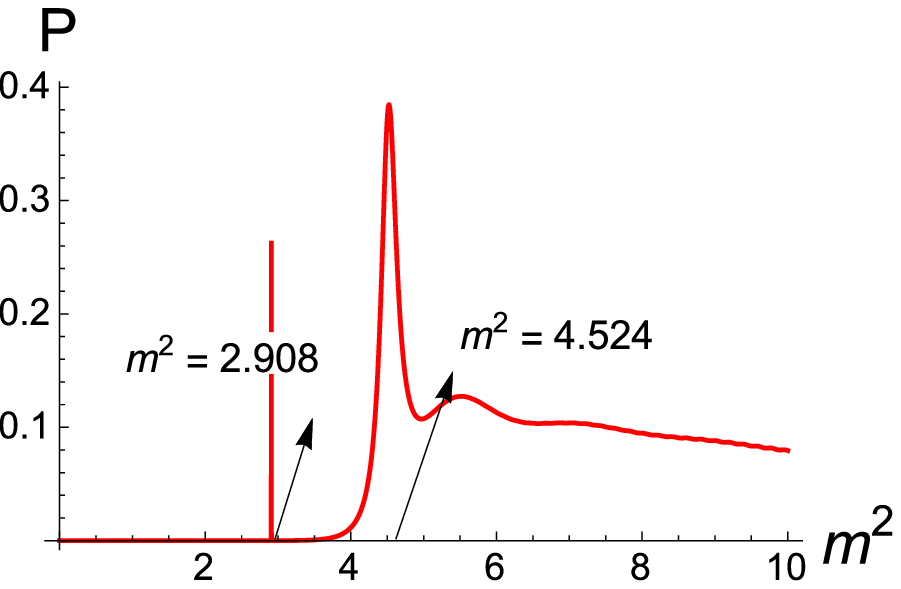}}
\subfigure[]{
\includegraphics[width=2.6in]{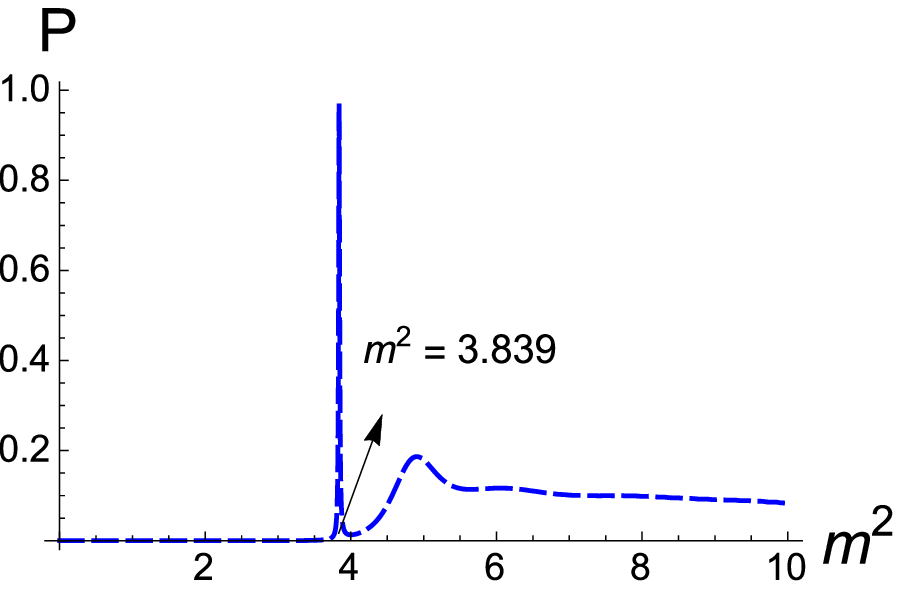}}
\end{center}
\caption{
Plots of the relative probability for even-parity or odd-parity KR field with the parameters $k=-21$ and $\beta=0.2$.}
\label{Relativeprobability1}
\end{figure}

\begin{figure}[htb]
\begin{center}
\subfigure[$m^2=2.908$]{
\includegraphics[width=2.6in]{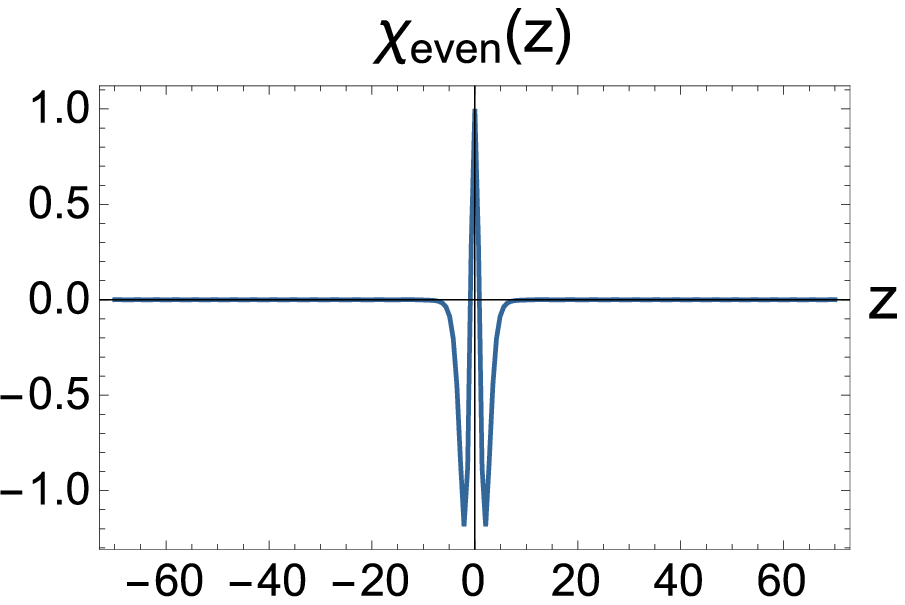}}
\subfigure[$m^2=4.524$]{
\includegraphics[width=2.6in]{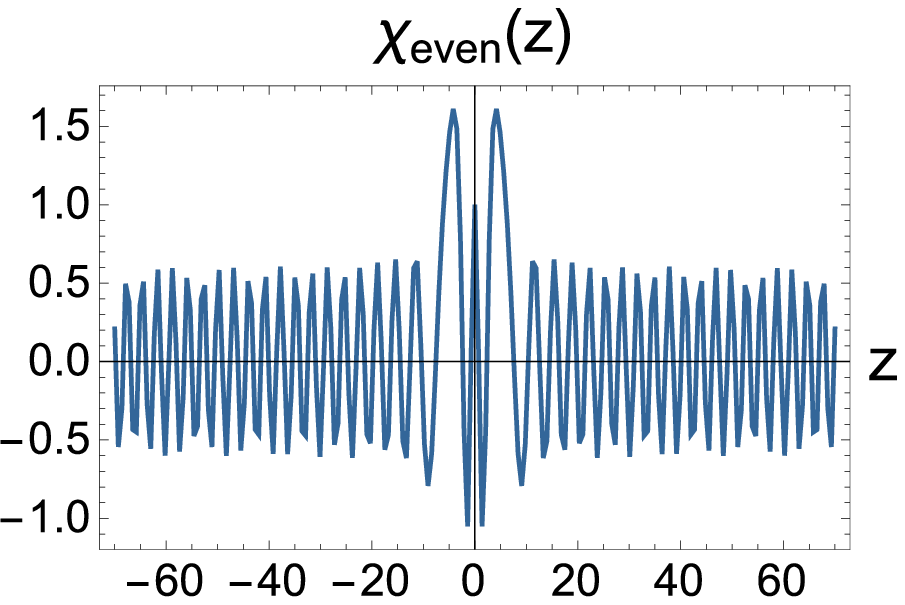}}
\subfigure[$m^2=3.839$]{
\includegraphics[width=2.6in]{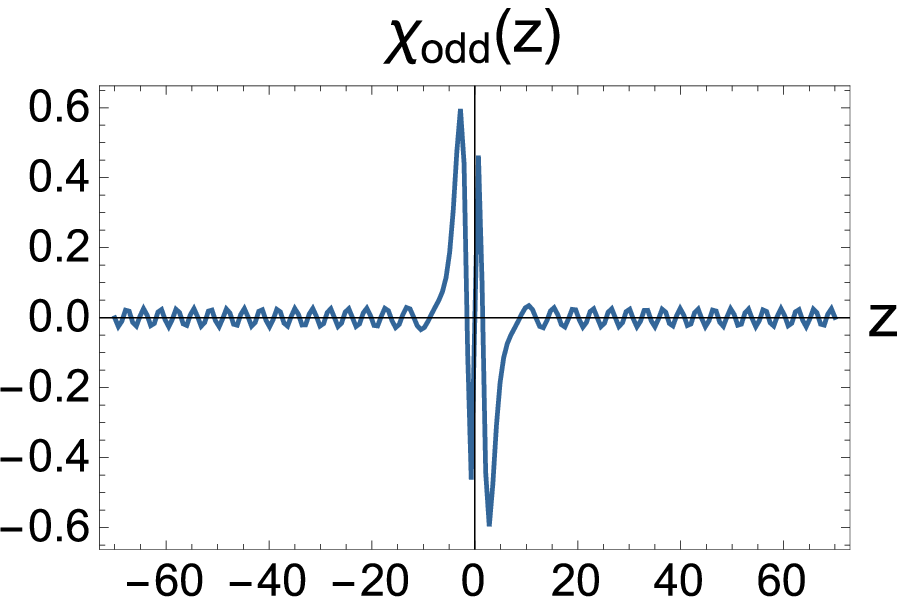}}
\end{center}
\caption{
Plots of the wave function  of even-parity or odd-parity resonances with  $k=-21$ and $\beta=0.2$.}
\label{Rwavefunction}
\end{figure}

Here we choose a relatively complicated coupling  $f(\phi)$ with negative parameter $k$.
 From Eq. (\ref{firstderivative}), the function $\Omega(z)$ for case III reads
 \begin{equation}
 \Omega(z)=e^{-\frac{7 A(z)}{2}} \text{sech}^{-\frac{k}{2}}\left(\frac{a z}{b}\right)
 \end{equation}
From the expression of effective potential in Eq. (\ref{effectvepotential}),  we can  get the effective potential of the Schr\"{o}dinger-like equation as follows:
\begin{equation}
V_
{\text{eff}}(z)=\left(4 A'(z)+\frac{\Omega'(z)}{\Omega(z)}\right)^2-\left(4 A'(z)+\frac{\Omega'(z)}{\Omega(z)}\right)'=\frac{\beta ^2 k \left(\beta ^2 (k-1) z^2+1\right)}{\left(\beta ^2 z^2+1\right)^2}.
\end{equation}
Note that it is independant of the brane thickness parameter $\delta$. The shape of effective potential is  plotted in figure (\ref{Volcanowavefunction}).
 As the figure shown, with the increase of $\beta$ or the decrease of $k$,
the potential well becomes thinner and deeper. The potential is a modified volcano-type potential  since it has the following asymptotic behaviors: it tends to zero at $z\rightarrow\infty$, and at $z=0$, the potential reaches its minimum $\beta^{2}k$ when $k<0$. For this type of volcano-like potential, there is no mass gap to separate zero mode from the KK massive modes, but there may exist some resonant modes  of  KR field.
For the zero mode, the normalization condition is
\begin{equation}
\int_{-\infty}^{+\infty}\left(1+z^{2} \beta^{2}\right)^{k} d z<\infty,
\end{equation}
which is finite for $k<-\frac{1}{2}$. So the zero mode  can be localized on the brane under the condition $k<-\frac{1}{2}$.
 In the following,  we  use the numerical method given in Refs.~\cite{Liu2009a,Liu2009c,Du2013} to find the resonant states of
 KR field for this potential, where the authors defined a relative probability, and the large relative probability $P$ for massive KK modes indicates the existence of resonances. Here, for instance, we set the parameters as $\beta=0.2$, $k=-21$  and find two or one resonant state. When the wave functions are either even-parity or odd-parity, the shapes of the relative probability are plotted respectively  in Fig.~\ref{Relativeprobability1},
where the masses of resonances are marked as $m^2=2.908,4.524$ or $m^2=3.839$. Correspondingly,  the  wave function  of resonances for even-parity or odd-parity KR field are plotted in Fig. ~\ref{Rwavefunction}.

\begin{figure}[htb]
\begin{center}
\subfigure[$\beta=0.32$, $k=-21$]{
\includegraphics[width=2.6in]{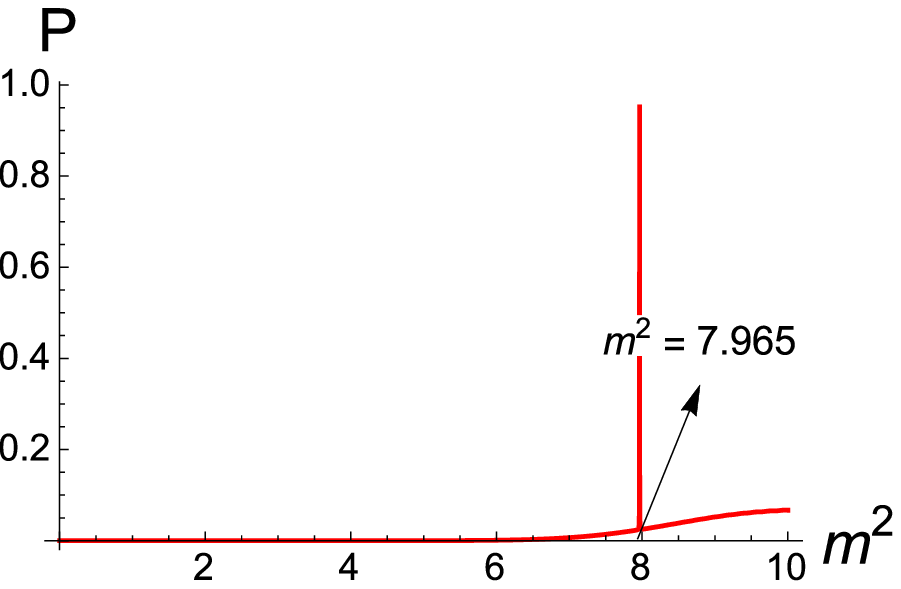}}
\subfigure[$\beta=0.2$, $k=-15$]{
\includegraphics[width=2.6in]{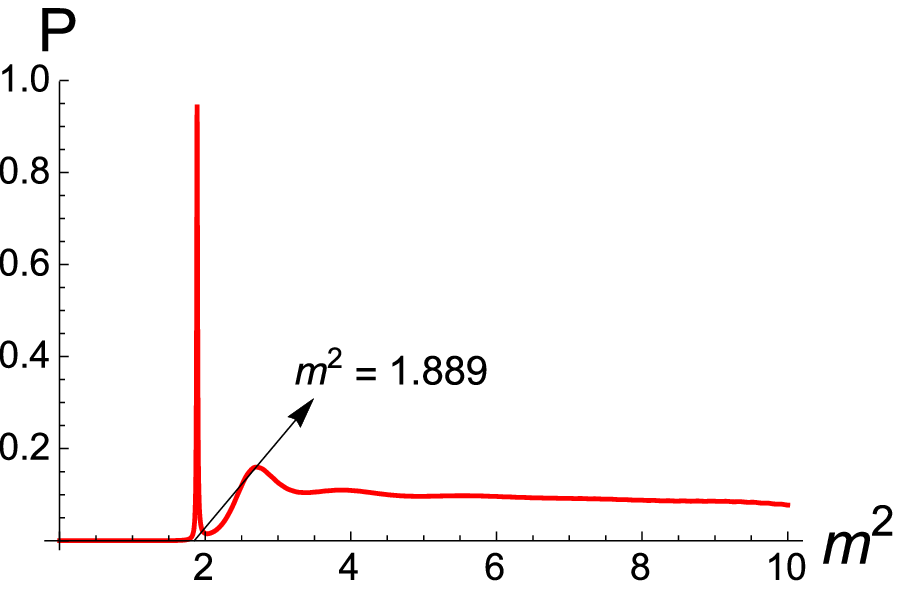}}
\end{center}
\caption{
Plots of the relative probability of resonances for even-parity KR field with different parameters $k$ and $\beta$.}
\label{Relativeprobability2}
\end{figure}

 If we choose another two sets of parameters as  $\beta=0.32$, $k=-21$, or $\beta=0.2$, $k=-15$, the shapes of the relative probability for even-parity state are plotted in Fig.~\ref{Relativeprobability2}.
Comparing Fig.\ref{Relativeprobability1} with  Fig.\ref{Relativeprobability2} , we find the  mass of the first resonant $KK$  mode increases with $\beta$ and the absolute value of $k$ (i.e., $|k|$). Here, we just list the mass spectra for the even-parity resonant state in Table ~\ref{tableA}.

\begin{table*}
\begin{ruledtabular}
\centering
\caption{The masses of  first even-parity resonance in relation to
 $\beta$ and $k$.}
\label{tableA}


 \begin{tabular}{|c|c|c|c|}

$\beta\;($k=-21$)$   & $m^{2}$    & $k\;(\beta=0.2)$
 & $m^{2}$   \tabularnewline
\hline
0.2  & 1.889   & -15  & 1.889 \tabularnewline
\hline
0.32  & 4.999   & -21  & 2.908  \tabularnewline
\hline
0.4  & 8.025    & -25  & 5.944  \tabularnewline
\hline
0.5 & 12.067   & -30  & 7.742   \tabularnewline
\end{tabular}
\vspace*{0cm}  
\end{ruledtabular}
\end{table*}

\section{Conclusion}\label{conclusion}
In this letter we investigate the localization of Kalb-Ramond gauge field on a non-flat  dS brane.  There are three important parameters  $\beta$, $\delta$ and $k$ in our model, which describe the expanding of the dS brane,  the thickness of the brane, and  the power of the coupling function, respectively.  We mainly investigate the effects of these parameters on the localization  and seek for the localization mechanism for KR field.
   \\ \indent The first coupling $\phi^{k}$ does not guarantee the localization of KR tensorial zero modes. The reason for such a behavior is that  the background solution of the parameter $\delta$ exactly breaks the localization condition. Secondly, by choosing the coupling $cos^{k}\phi$, the effective potential is  PT-like potential influenced by three parameters $\beta$, $\eta$ and $k$, and the KR field zero mode can be localized on the brane under the condition  $k > \delta$.  With the increase of $\beta$ and $k$, or the decrease of $\delta$,  the
effective potential becomes deeper and thinner.   A set of analytic solutions for the  mass spectra and wave function for the KK modes are obtained. Finally, for the third case, the effective potential  with negative $k$ is a volcano-like one which indicates the possibility of appearance of resonant modes. The zero mode of KR field can be localized on this thick brane under the condition $k<-1/2$.  The numerical analysis of resonant state reveals that the volcano-like potential is affected  by  only two parameters $\beta$ and $k$, and it becomes thinner and deeper with the increase of $\beta$ and $|k|$. As is shown in Table~\ref{tableA},  the  mass of the first resonant mode also increases with the values of $\beta$ and  $|k|$.

\section*{Acknowledgement}
C. Yang sincerely thanks Drs. Tao-Tao Sui, Qin-Tan, and Zheng-Quan Cui for useful discussions. This work was supported by the National Natural Science Foundation of China (Grants No. 11705070) and the Fundamental Research Fund for Physics of Lanzhou University (Grants No. lzujbky-2019-ct06).


\begin{thebibliography}{10}

\bibitem{Arkani-Hamed1998}
 N. Arkani-Hamed, S. Dimopoulos, and G. R. Dvali,
 Phys. Lett. {\bf B 429}, 263 (1998).

\bibitem{Randall1999}
 L. Randall and R. Sundrum,
 Phys. Rev. Lett. {\bf 83}, 3370 (1999).




\bibitem{Guo2015b}
 B. Guo, Y.-X. Liu, K. Yang, and X.-H. Meng,


\bibitem{Yang2012}
 K. Yang, Y.-X. Liu, Y. Zhong, X.-L. Du, and S.-W. Wei,
 Phys. Rev. {\bf D 86}, 127502 (2012). 


\bibitem{Gogberashvili2002}
 M. Gogberashvili,
 Int. J. Mod. Phys. {\bf D 11}, 163 (2002).


\bibitem{Antoniadis1998}
 I. Antoniadis, N. Arkani-Hamed, S. Dimopoulos, and G. R. Dvali,
  Phys. Lett. {\bf B 436}, 257 (1998).



\bibitem{Arkani-Hamed2000}
 N. Arkani-Hamed, S. Dimopoulos, N. Kaloper, and R. Sundrum,
 Phys. Lett. {\bf B 480}, 193 (2000).


\bibitem{Neupane2011}
 I. P. Neupane,
 Phys. Rev. {\bf D 83}, 086004 (2011).





\bibitem{Starkman2001}
 G. D. Starkman, D. Stojkovic, and M. Trodden,
 Phys. Rev. Lett. {\bf 87}, 231303 (2001).





\bibitem{Kim2001}
 J. E. Kim, B. Kyae, and H. M. Lee,
 Phys. Rev. Lett. {\bf 86} , 4223 (2001).


\bibitem{Randall1999b}
 L. Randall and R. Sundrum,
  Phys. Rev. Lett. {\bf 83}, 4690 (1999).

\bibitem{Rubakov:1983rsd}
 V. A. Rubakov and M. E. Shaposhnikov,   {Phys. Lett. \textbf {B 125} (1983) 136}.



\bibitem{Chumbes2012}
A. E. R. Chumbes, J. M. Hoff da Silva, and M. B. Hott,
 Phys. Rev. {\bf D 85}, 085003 (2012).



 \bibitem{BajcPLB2000}
A. D. Furlong, A. H. Aguilar, R. Linares, R. R. M. Luna, and H. A. M. Tecotl,
Gen. Rel. Grav. \textbf{46}, 1815 (2014).

\bibitem{Liu0708}
Y.-X. Liu, X.-H. Zhang, L.-D. Zhang, and Y.-S. Duan,
JHEP \textbf{0802}, 067 (2008).

\bibitem{Koroteev08}
R. Casana, A. R. Gomes, and F. C. Simas,
JHEP \textbf{1506}, 135 (2015).

\bibitem{Flachi09}
A. Flachi and M. Minamitsuji,
Phys. Rev. \textbf{D 79}, 104021 (2009).

\bibitem{Neronov2001}
A. Neronov,
  {Phys. Rev.} {\bf D 64}, 044018 (2001).



\bibitem{Christiansen2012}
H. Christiansen and M. Cunha,
   {Eur. Phys. J.} {\bf C 72}, 1942 (2012).


\bibitem{Cruz2009}
W. T. Cruz, M. O. Tahim, and C. A. S. Almeida, Europhys. Lett. {\bf 88}, 41001 (2009).

\bibitem{Christiansen2010} H. Christiansen, M. Cunha, and M. Tahim, Phys. Rev. {\bf D 82 }, 085023 (2010).


\bibitem{Cruz2013} W. Cruz, R. Maluf, and C. Almeida, Eur. Phys. J. {\bf C 73 }, 2523    (2013).


 \bibitem{Tahim2009}
M. O. Tahim, W. T. Cruz, and C. Almeida,
   {Phys. Rev.} {\bf D 79}, 085022 (2009).






\bibitem{ThickBrane2}
N. B. Cendejas, D. M. Morejon, and R. R. M. Luna,
Gen. Rel. Grav. \textbf{47}, 77 (2015).


\bibitem{LiuXu2014}
 Y.-X. Liu, Z.-G. Xu, F.-W. Chen, and S.-W. Wei,
 Phys. Rev. \textbf{D 89}, 086001 (2014).




\bibitem{LiYY2017}
Y.-Y. Li, Y.-P. Zhang, W.-D. Guo, and Y.-X. Liu, Phys. Rev. {\bf D 95}, 115003 (2017)


\bibitem{Davoudias2000}
 H. Davoudias, J. L. Hewett, and T. G. Rizzo, Phys. Lett. {\bf B 473}, 43 (2000).

\bibitem{Liu2008} Y.-X. Liu, S.-W. Wei,  and Y.-S. Duan,
JHEP {\bf 08}, 041 (2008).


\bibitem{Oda2000} I. Oda, Phys. Lett. {\bf B 496}, 113 (2000).


\bibitem{Kehagias2001}
A. Kehagias and K. Tamvakis,  Phys. Lett. {\bf B 504}, 38 (2001).

\bibitem{Vaquera-Araujo2015}
C. A. Vaquera-Araujo and O. Corradini,  Eur. Phys. J. {\bf C 75}, 48(2015).


\bibitem{Odd2001}
I. Oda, A new mechanism for trapping of photon.

\bibitem{Zhaozh2017}
 Z.-H. Zhao and  Q.-Y. Xie, JHEP {\bf 05}, 072 (2018).



\bibitem{Green1985}
M. Green, J. Schwarz, and E. Witten, Superstring Theory vol II (Cambridge, 1985).


\bibitem{Vasilic2008}
M. Vasilic and M. Vojinovic, Phys. Rev. {\bf D 78}, (2008) 104002.



\bibitem{Moazzen2017}
M. Moazzen, Int. J. Mod. Phys. {\bf A  32},  1750058 (2017).


\bibitem{Goetz1990}
G. Goetz, J. Math. Phys. {\bf 31}, 2683 (1990).

\bibitem{Gass1999}
 R. Gass and M. Mukherjee, Phys. Rev. {\bf D 60}, 065011 (1999).



\bibitem{Liu2009a}
Y.-X. Liu, C.-E. Fu, L. Zhao, and Y.-S. Duan,
    {Phys. Rev.} {\bf D 80}, 065020 (2009).

\bibitem{Liu2009c}
Y.-X. Liu, J. Yang, Z.-H. Zhao, C.-E. Fu, and Y.-S. Duan,
  {Phys. Rev.} {\bf D 80}, 065019 (2009).



\bibitem{Du2013}
Y.-Z. Du, L. Zhao, Y. Zhong, C.-E. Fu, and H. Guo,
  {Phys. Rev.} {\bf D 88}, 024009 (2009).



\end{thebibliography}
\end{document}